\documentclass[runningheads]{llncs}
\usepackage[T1]{fontenc}
\usepackage{amsmath}
\usepackage{graphicx}
\usepackage[table,xcdraw]{xcolor}
\usepackage[font=small,skip=2pt]{caption}
\usepackage[skip=0pt]{subcaption}
\usepackage{multicol}
\usepackage{multirow}
\usepackage{adjustbox}
\usepackage{hyperref}
\usepackage{color}
\usepackage{float}
\usepackage{lipsum}
\usepackage[utf8]{inputenc}
\usepackage{subcaption}
\usepackage{placeins}

\begin{document}
\title{A Comparison of Baseline Models and a Transformer Network for SOC Prediction in Lithium-Ion Batteries}

\titlerunning{A Comparison of Machine Learning Models for SOC Prediction} 
\author{Hadeel Aboueidah\orcidID{0009-0005-6385-7444} and
Abdulrahman Altahhan\orcidID{0000-0003-1133-7744}}
\institute{University of Leeds, School of Computer Science, Leeds, United Kingdom. \\
\email{hadeel.ellian@gmail.com, a.altahhan@leeds.ac.uk}}

\maketitle
\bibliographystyle{splncs04}
\begin{abstract}

    Accurately predicting the state of charge of Lithium-ion batteries is essential to the performance of battery management systems of electric vehicles. One of the main reasons for the slow global adoption of electric cars is driving range anxiety. The ability of a battery management system to accurately estimate the state of charge can help alleviate this problem. In this paper, a comparison between data-driven state-of-charge estimation methods is conducted. The paper compares various neural network-based models and common regression models for SOC estimation. These models include several ablated transformer networks, a neural network, a lasso regression model, a linear regression model and a decision tree. Results of various experiments conducted on data obtained from natural driving cycles of the  BMW i3 battery show that the decision tree outperformed all other models including the more complex transformer network with self-attention and positional encoding. The decision tree model scored perfect values of performance metrics (MSE=0, $R^2$=1, RMSE=0, MAE=0). This paper could be helpful for researchers to select appropriate data-driven methods for SOC estimation of lithium-ion batteries in real driving cycles.  

\keywords{Transformer, machine learning, deep learning, linear regression, Li-ion battery, state of charge, electric vehicle}
\end{abstract}

\section{Introduction and Background Research }

 With the growing concern about climate change and global warming, the use of electric vehicles (EVs)  has emerged as a sustainable solution and an environmentally friendly alternative to petrol-fueled vehicles. Zero-emissions vehicles have become an essential part of every country's futuristic and sustainable vision. Rechargeable lithium-ion batteries are widely used for EVs and are considered the best available option, offering suitable battery energy density and cycle life  \cite{LION}.  Moreover, the development of (Li-ion) batteries over the next years will lead to more complex battery dynamics as well as higher energy density. Therefore, advancement in battery management systems (BMS) that can optimize and monitor battery behaviour will be required in parallel with the development of  Li-ion batteries and the overall electrification system \cite{DataDrivenHealth}. 
 
 The state of charge (SOC) is one of the most important parameters in a battery management system (BMS).  It is a metric that provides information about the battery's remaining capacity under  current working conditions. Unlike the fuel gauge in a petrol-fueled vehicle, SOC cannot be directly measured in EVs. Therefore, it needs to be estimated in practical applications which can be a challenging task  \cite{SOCDefintion}  The equation of a battery's SOC is defined as follows :
 \begin{equation}
SOC =  \frac{C_{curr}}{C_{full}} \times 100 \% ,
\end{equation}
where $C_{curr}$ is the capacity of the battery in its current state and $C_{full}$ is the
the capacity of the battery when it is fully charged. 

The conventional SOC estimation methods include open-circuit voltage, impedance-based estimation, fuzzy logic, Kalman filter, and model-based estimation. The model-based methods have an advantage over the other methods as they can be used for online applications. Moreover, physics-based models (PBMs) are an advancement of model-based methods. The pseudo-two-dimensional (P2D) is the most studied PBM model in the literature. This model provides us with a better understanding of the battery's internal dynamics such as the concentration of lithium-ion in the electrodes and electrolytes. However, it is less practical for online applications due to its complex governing equations that require higher computational costs. Additionally, traditional PBMs do not consider the material properties within the battery, which are essential for understanding the battery’s degradation behaviour and its SOC.  The main battery models used today for real-time SOC estimation in BMSs are equivalent circuit models (ECMs).   These models are widely used because of their low computational cost; however, one drawback is that their accuracy is limited when predicting battery characteristics across a range of varying operating conditions \cite{ConventionalSOC}\cite{SOCMethodsReview}.  Fig \ref{common_SOC_estimation_methods} below demonstrates the model's Accuracy Vs CPU time for the most studied Li-ion battery models in the literature, ECMs and PBMs :
  \begin{figure} [H]
        \centering
        \includegraphics[width= 0.6 \textwidth] {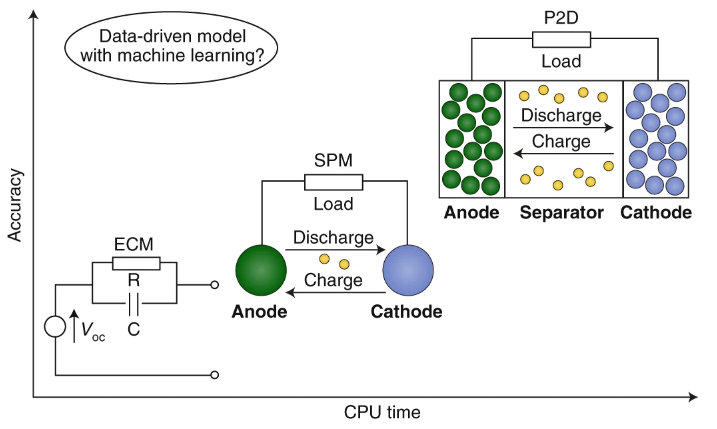}
        \caption{Accuracy vs. CPU time for conventional SOC estimation methods  }
        \label{common SOC estimation methods}
    \end{figure}
As shown in Fig. \ref{common SOC estimation methods}\cite{ConventionalSOC} above,  the ECM SOC estimation model requires less CPU time but at the expense of accuracy. On the other hand, advanced PBM models, such as P2D, offer higher accuracy but come with greater computational cost and time. This trade-off in the performance of these two battery models prompts the investigation of data-driven methods (DDMs) for SOC prediction in Li-ion batteries. 

Oftentimes, researchers aim to predict the future behaviour of a battery such as the remaining range an EV can drive. Hence, What if we had a function that could take the current state of a certain battery characteristic and output its future behavior without relying on detailed physical knowledge? Fortunately, this can be achieved using data-driven methods (DDMs), such as baseline machine learning and neural network-based models. In the following pages, we will explore the advantages of utilizing these methods to predict the SOC of Li-ion batteries, followed by an examination of their use in the literature. 

Moreover, the use of data-driven methods (DDM) to predict the state of charge of lithium-ion batteries in electric vehicles has several advantages such as driving range estimation. Determining the range of an EV depends mainly on the battery’s SOC and is crucial for trip planning and addressing unexpected battery depletion \cite{StationRange}.  Another advantage of using DDMs for SOC prediction is battery life optimization. Since EV batteries have a limited lifespan, proper maintenance is essential to ensure maximum endurance. By predicting the battery’s state of charge, charging and discharging cycles can be optimized, thus extending the battery’s life and reducing overall costs \cite{BatteryLife}. In addition, data-driven SOC prediction can aid in planning charging station infrastructure by analyzing charging patterns to determine optimal locations for stations. Consequently, this will alleviate driving range anxiety and promote the adoption of electric vehicles \cite{chargeStation} \cite{Station&Range}. Another reason why predicting SOC using DDMs is beneficial is its potential to enhance vehicle-to-grid (V2G) services. These services allow EVs to function as power sources, discharging battery energy back into the grid during periods of peak demand. Therefore, an accurate prediction of an EV battery’s SOC is vital for V2G services, as it ensures that the vehicle has sufficient charge to complete the trip while still supplying the grid with the necessary power  \cite{V2G}. 

Furthermore, the use of data-driven methods for SOC prediction of lithium-ion batteries has surged in recent years and has been widely explored in the literature. In \cite{TurkishPaper}, a bagging random forest model was proposed to predict the SOC of the Li-ion batteries. The model was trained on 32067 instances of the dataset we are utilising in this paper.  The proposed model achieved an MAE value of 0.280 and an RMSE value of 0.519, which are relatively good results. These values will be used as a benchmark to evaluate the performance of the machine learning models applied in our study. 

The task of SOC prediction has been explored using various datasets and architectures. Neural network-based architectures were chosen due to their high accuracy and ability to handle data-rich systems, as demonstrated in studies such as \cite{nn1}\cite{nn2} and \cite{nn3}. Complex neural network models, such as transformers, have also been used for SOC regression. In \cite{Trans1}, a transformer network with an L1 robust observer was employed to estimate SOC in EVs, showing desirable low MAE and RMSE values across varying temperatures. Another study \cite{Trans2} proposed a transformer with an immersion and invariance adaptive observer, achieving lower MAE and RMSE values (0.49-1.12 for MAE and 0.54-1.39 for RMSE) compared to a baseline transformer and LSTM.  Given their widespread use in the literature, the performance of transformers will be investigated using the data used in our SOC prediction task. 

Even though decision trees are known for effectively capturing non-linear patterns, it was observed that they are seldom selected for SOC prediction of Li-ion batteries in the literature, as more complex neural network-based models are usually sought after. 

In the following sections, we will explore the potential of various baseline ML methods, as well as neural network-based methods, in the methodology section. Then, we will compare and analyze the performance of these models in the experimental results and discussion section. Finally, the main findings and conclusions drawn from this study will be presented in the conclusion section.  
 
\section{Methodology}

\subsection{The data}
To compare the performance of these different regression models, 736,983 records were obtained from 72 real driving trips with a BMW i3 (60 Ah). The numerical features were standardized using the StandardScaler function, and 24 features were selected from a total of 51. Additionally, the confusion matrix obtained during the exploratory data analysis (EDA) highlighted the non-linear relationship between the input features and the target. The correlation with the target and the variance of each selected feature across the dataset were considered during the feature selection process. Moreover, 78,657 outliers were removed from the dataset after calculating the z-score and setting a threshold of 3 to identify outliers. The cleaned dataset was then used for further analysis. The original dataset can be accessed via this link: https://ieee-dataport.org/open-access/battery-and-heating-data-real-driving-cycles \cite{data}. 

\subsection{Evaluation metrics}
The three evaluation metrics used to evaluate the models’ performance in this SOC prediction task are: Mean Squared Error (MSE), Root Mean Square Error (RMSE), R², and Mean Absolute Error (MAE). The equations for these evaluation metrics are illustrated below: 
\begin{equation}
MSE = \frac{1}{n}\sum_{i=1}^{n}(y_i - \hat{y_i})^2
\end{equation}
\begin{equation}
RMSE = \sqrt{\frac{1}{n}\sum_{i=1}^{n}(y_i - \hat{y_i})^2}
\end{equation}
\begin{equation}
R^2 = 1 - \frac{\sum_{i=1}^{n}(y_i - \hat{y_i})^2}{\sum_{i=1}^{n}(y_i - \bar{y})^2}
\end{equation}
.
\begin{equation}
MAE = \frac{1}{n} \sum_{i=1}^{n} \left| y_i - \hat{y}_i \right|
\end{equation}

, where n is the number of samples, ${y_i}$ is the actual value of the target variable for the $i^{th}$ sample, 
 $\bar{y_i}$ is the mean of the target variable, and $\hat{y_i}$  is the predicted value of the target variable for the $i^{th}$ sample.

 \subsection{Baseline approaches}
In this paper, various machine learning and neural network-based models were tested to predict the state of charge of lithium-ion batteries from historical natural driving cycles. These models included several scikit-learn models, such as linear regression, lasso regression, and decision tree, as well as neural network-based models from Keras, such as neural network regression and the transformer network. These models are listed and illustrated below:
   
 \subsubsection {Linear Regression}
The first baseline regression model used to predict the lithium battery SOC is linear regression from scikit-learn. In linear regression, a set of input features is represented by the variable X and matched with its corresponding continuous output values y.  The linear regression model is performed through a straight-line approximation of the outputs \cite{Theorylinearregression}.  Its equation is given below:

\begin{equation}
    Y_i  = \beta_0 + \beta_1 X_i +  \epsilon_i  
\end{equation}
    
, where Y is the dependent variable, X is the independent variable,  $\beta_0 $  is the intercept of the line, $\beta_1 $  is the linear regression coefficient (slope of the line), and $ \epsilon $  is the random error term. 

\subsubsection {Lasso Regression} 
In lasso regression, which stands for least absolute shrinkage and selection operator, linear regression is performed, but with a reduced number of features. This is achieved using L1 regularization, where a penalty term is added to the cost function to encourage the model to select only the most important attributes and set the coefficients of less important attributes to zero.\cite{TheoryLasso}.  The lasso regression equation can be given as follows: 
\begin{equation}
\hat{\beta}^{lasso} = \underset{\beta}{\operatorname{argmin}} \left\lbrace \frac{1}{2n} \sum_{i=1}^{n}(y_i - \beta_0 - \sum_{j=1}^{p}x_{ij}\beta_j)^2 + \lambda \sum_{j=1}^{p}|\beta_j| \right\rbrace
\end{equation}

, where  $\hat{\beta}^{lasso}$ represents the estimated regression coefficients for  Lasso regression, $n$ and $p$ represent the number of observations and predictor variables, respectively. $\beta_0$ is the intercept term, $y_i$ is the $i$th observation of the target variable and $x_{ij}$ is the $ij$th observation of the $j$th predictor variable.

   \subsubsection {Decision Tree} 
 A decision tree regression can be highly effective when the data cannot be modeled with a straight line, as is the case with the dataset used in this paper. Hence, the potential of decision tree regression was explored for this SOC prediction task.  
 
  \subsubsection {Neural network regression}  
Neural network regression for SOC values was performed using the Keras API in TensorFlow. The network used for this task consists of five layers: one input layer, three hidden layers, and one output layer. The input layer has the same shape as the number of selected features. The model includes three dense layers with 32, 8, and 2 neurons, respectively, each using the rectified linear unit (ReLU) activation function, whose equation is given below: 
\begin{align*}
   \label{eq:sum_i}
   \centering
   Relu(z) = max(0, z)
   \end{align*}

The ReLU activation function was chosen because it is easier to compute and faster to evaluate, as it requires only a single comparison operator. It can also  tackle the vanishing gradient problem and allow the model to learn from large datasets. \cite{ReLU} 
   
Moreover, the output layer consists of a single neuron without an activation function. The model was compiled using the Adam optimizer with a learning rate of 0.001 and mean squared error (MSE) as the loss function. The network was then trained on the training set (X train, y train) for 100 epochs with a batch size of 1000.

\subsubsection {Optimisation of the four baseline models} 
The hyperparameters for the baseline models were set as follows: an empty dictionary for linear regression, alpha values for Lasso regression, and max depth for the decision tree. These hyperparameters were optimized using GridSearch for each model. K-fold cross-validation was applied to evaluate performance and prevent overfitting. For the baseline neural network regression, three hidden layers were added to ensure efficient performance and minimal training time, with early stopping employed to halt training if the loss did not improve after 5 epochs. Performance metrics, including MSE, RMSE, MAE, and $R^2$ were calculated to compare the models. 
   \vspace {-0.1 cm}
   
\subsection{The Transformer Network and its ablated versions }
 
\begin{figure}
    \centering
    \begin{subfigure}[b]{0.4\textwidth} 
        \includegraphics[width=\textwidth]{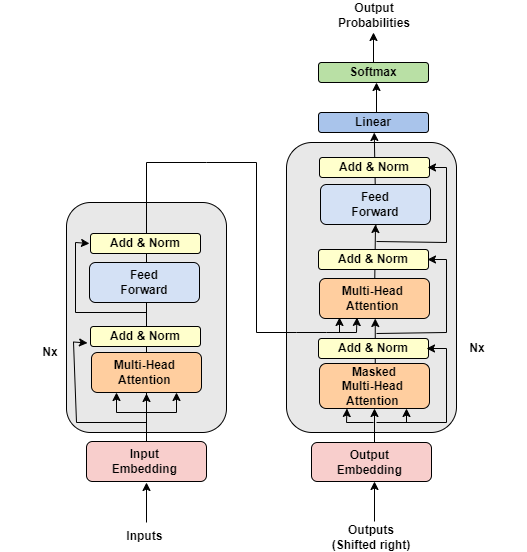}
        \caption{The transformer with self-attention only.}
        \label{fig:figure1}
    \end{subfigure}
    \hspace{1cm} 
    \begin{subfigure}[b]{0.4\textwidth} 
        \includegraphics[width=\textwidth]{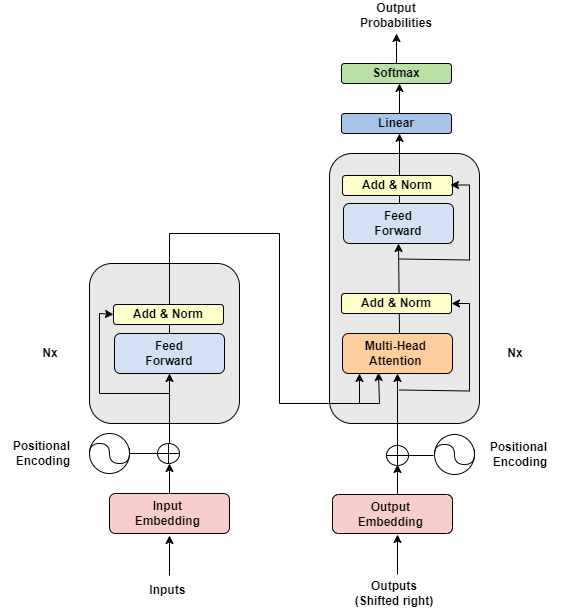}
        \caption{The transformer with positional encoding only.}
        \label{Transformer with P }
    \end{subfigure}
    \caption{The two ablated transformer networks}
    \label{Transformer with A}
\end{figure}

 The potential of using a transformer network to predict the state of charge of lithium-ion batteries was explored. The transformer network was ablated with some of its main parts removed each time to assess the necessity of these parts and determine if a less complex architecture could outperform the original model in this regression task.  The two diagrams of the ablated versions of the transformer network are demonstrated in fig.\ref{Transformer with A}. 

As shown in fig.\ref{Transformer with A} , the self-attention and positional encoding are ablated each time to create less complex networks. The significance of these two layers will be further investigated by evaluating the performance of the ablated transformer networks in the following sections. 

\subsubsection {Transformer with Self-Attention Only}

A transformer network architecture was built using the Keras functional API. The input layer was reshaped to (batch size, 1, input shape) for compatibility. The model architecture included a multi-head attention layer, dropout, layer normalization, two fully connected layers, and an output layer without an activation function. The Adam optimizer, with a learning rate of 0.001, was used alongside mean squared error as the loss function. Early stopping was implemented to prevent overfitting and reduce training time. The model was then trained on the training data and evaluated using the validation data.  

\subsubsection {Transformer with Positional encoding Only}

 In this transformer network, a positional encoding layer was utilized, while the attention layer was removed to simplify the architecture. Constants and hyperparameters such as embedding size, dropout rate, hidden units, and batch size were defined. The input layer was reshaped to add an extra dimension, followed by the addition of a custom positional encoding layer with dropout. Layer normalization was applied to the combined output, which was then passed through two dense layers with ReLU activation. Finally, a single output layer was added to predict the target value. The model was compiled using the Adam optimizer and mean squared error as the loss function, then trained and validated.  

\subsubsection {Transformer with Self-Attention and Positional Encoding }

This model follows the traditional transformer architecture with self-attention and positional encoding, as introduced in "Attention is All You Need"\cite{attention}. The input layer corresponds to the number of features in the training dataset and is reshaped into a 3D tensor. Positional encoding is applied using an embedding layer, with dropout added to prevent overfitting. A multi-head attention layer is included to capture dependencies between input features, followed by a skip connection and layer normalization. A dense layer with ReLU activation is then applied, and the final output is generated through a dense layer with a linear activation function. The model was compiled and trained with a batch size of 1000 for 100 epochs, and its performance was evaluated using MSE on the validation set. 

\section{Experiment Results and Discussions}
    
The baseline models and transformer networks were trained on the dataset, and their performances were compared.
 
\begin{figure}[htbp]
    \centering
    \begin{subfigure}[b]{0.4\textwidth} 
        \includegraphics[width=\textwidth]{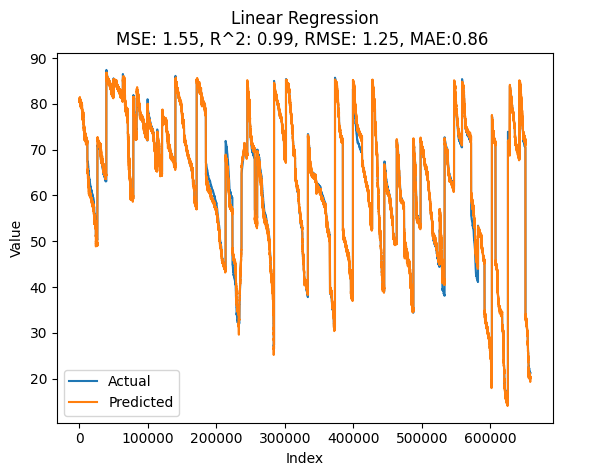}
        \caption{Predicted vs. actual SOC values for the linear regression model}
        \label{LR}
    \end{subfigure}
    \hfill
    \begin{subfigure}[b]{0.4\textwidth} 
        \includegraphics[width=\textwidth]{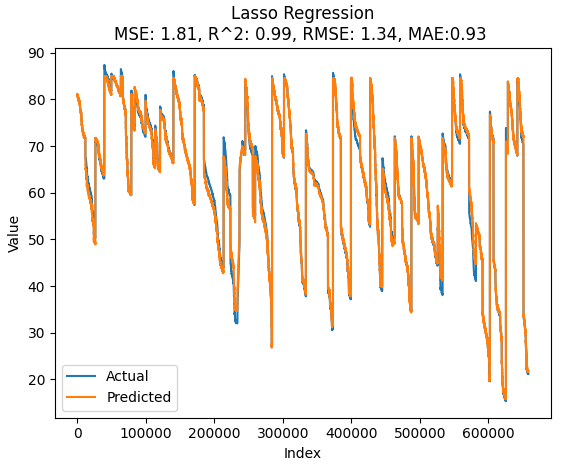}
        \caption{Predicted vs. actual SOC values for the lasso regression model}
        \label{lasso regression}
    \end{subfigure}
    \vspace{0.1cm} 

    \begin{subfigure}[b]{0.4\textwidth} 
        \includegraphics[width=\textwidth]{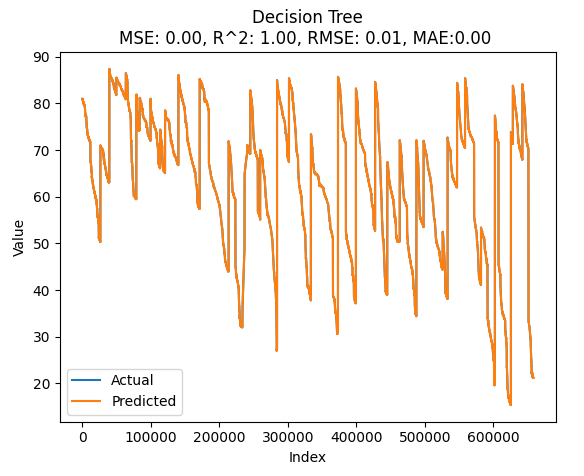}
        \caption{Predicted vs. actual SOC values for the decision tree model}
        \label{Decision tree}
    \end{subfigure}
    \hfill
    \begin{subfigure}[b]{0.4\textwidth} 
        \includegraphics[width=\textwidth]{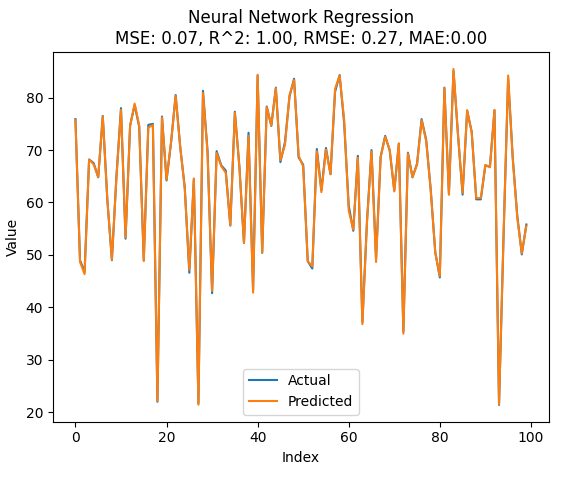}
        \caption{Predicted vs. actual SOC values for the NN model}
        \label{nn regression}
    \end{subfigure}
    \vspace{0.1cm} 

    \begin{subfigure}[b]{0.4\textwidth} 
        \includegraphics[width=\textwidth]{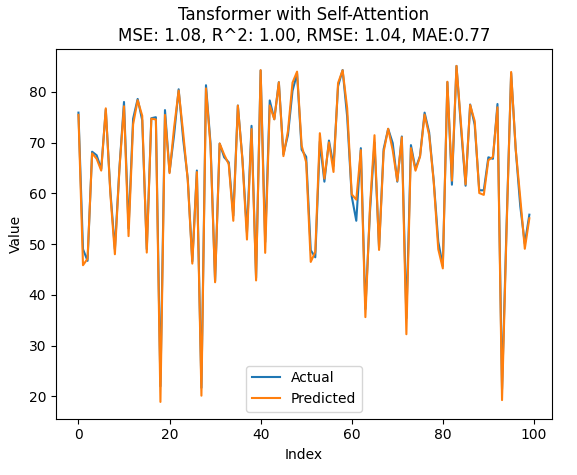}
        \caption{Predicted vs. actual SOC values for the transformer network with self-attention}
        \label{Transformer with self-attention}
    \end{subfigure}
    \hfill
    \begin{subfigure}[b]{0.4\textwidth} 
        \includegraphics[width=\textwidth]{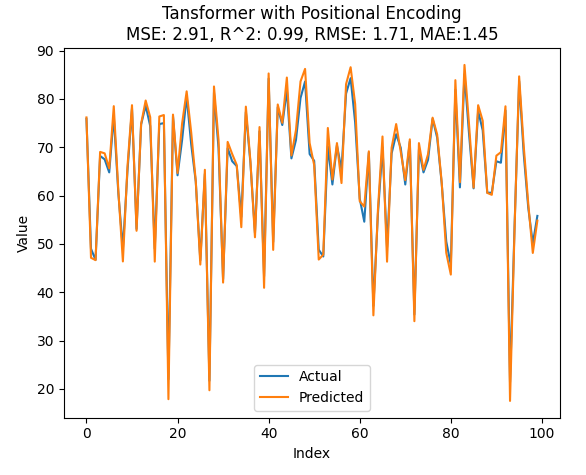}
        \caption{Predicted vs. actual SOC values for the transformer network with positional encoding}
        \label{Transformer with positional encoding}
    \end{subfigure}
    \vspace{0.1cm} 

    \begin{subfigure}[b]{0.4\textwidth} 
        \includegraphics[width=\textwidth]{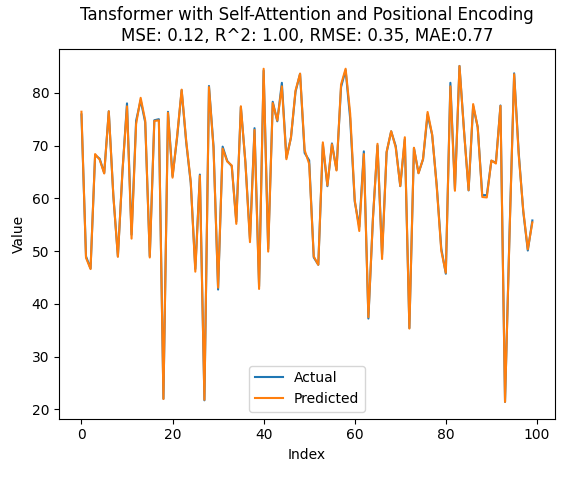}
        \caption{Predicted vs. actual SOC values for the transformer with self-attention and positional encoding}
        \label{Transformer with self-attention and positional encoding}
    \end{subfigure}
    \hfill
    \begin{subfigure}[b]{0.4\textwidth} 
        \includegraphics[width=\textwidth]{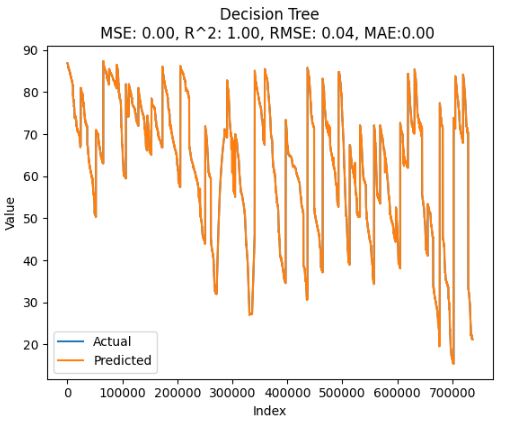}
        \caption{Predicted vs. actual SOC values for the decision tree model obtained from running the model on a smaller portion of the dataset}
        \label{Decision tree small}
    \end{subfigure}

    \caption{Predicted vs. actual SOC values for all models}
    \label{All models}
\end{figure}

The predicted vs actual SOC values for the LR model are demonstrated in  Fig. \ref{LR}. In the case of perfect prediction, the predicted graph would coincide with the actual graph. However, as seen in Fig. \ref{LR},  the predicted SOC graph deviates from the actual one, resulting in high MSE and $ R^2 $  values. Fig. \ref{lasso regression} highlights the lasso regression model's relatively poor performance on the validation set, also indicated by its high MSE and $ R^2 $  values.
In contrast, Fig. \ref{Decision tree} demonstrates that for the DT model, the predicted SOC graph closely matches the actual SOC graph, which is reflected in the near-perfect MSE and  $ R^2 $  values.  Fig.\ref{nn regression} shows that the neural network achieved a good MSE value of 0.07 and an RMSE value of 0.27.

Moving on to transformer networks, Fig.\ref{Transformer with self-attention} depicts the performance of the transformer network with only an attention layer.  The figure shows that the predicted SOC values for the transformer network with self-attention only slightly deviate across different data instances, with an MSE value of 1.08 and a perfect $ R^2$ value of 1. Fig.\ref{Transformer with positional encoding} presents the results for the \textit{transformer network with positional encoding only}. This network scored an MSE of 2.91, an RMSE of 1.71, and an almost perfect R² value of 0.99. Eventually, Fig.\ref{Transformer with self-attention and positional encoding} illustrates the performance of the conventional transformer network. As shown, this model performed competently, achieving an MSE of 0.12, an RMSE of 0.35, an MAE of 0.77, and a perfect R² value of 1. The predicted vs. actual SOC values for each model are demonstrated in fig.\ref{All models}.    

 \subsubsection {Overall performance}

After experimenting with all the baseline models and transformer-ablated architectures, the performance metrics indicate that the decision tree model outperformed not only all other baseline models but also the transformer networks tested in this prediction task. The decision tree achieved perfect performance metrics (MSE = 0, R² = 1, RMSE = 0, MAE = 0). This outstanding performance prompts us to explore the advantages the decision tree model over other architectures when predicting the real-life  SOC values of lithium-ion batteries. 
First, decision trees (DTs)  are simple and fast learners capable of capturing non-linear relationships between input variables (such as voltage, current, etc.) and the target variable (SOC) \cite{NonlinearRelationship} \cite{FeatureSelection}. Second, DTs are computationally efficient, making them ideal for real-time SOC prediction in battery management systems    \cite{FeatureSelection} . They are also robust to missing and noisy real-world data, which further explains their outstanding performance in this prediction task  \cite{Robustness}. 
Furthermore, the baseline decision tree model used in this task outperformed the proposed bagging random forest model in \cite {TurkishPaper} , which was trained and tested on a smaller dataset of only 32,067 instances with four time-dependent attributes. To ensure a fair comparison between the proposed model and the models used in our task, a sample of equal dimensions was extracted from our dataset and fed into the models investigated in this study.  Fig. \ref{Decision tree small}   presents the graph of the predicted vs. actual SOC values for the decision tree model, which was tested on a subset of the data that is equal in size to the input data used in \cite {TurkishPaper}. The baseline decision tree achieved an RMSE value of 0.4, along with perfect MSE and MAE values of 0, and an $R^2$ value of 1.  This model relatively outperformed the proposed bagging random forest in \cite {TurkishPaper}, which recorded an MAE of 0.280 and an RMSE of 0.519. The decision tree maintained its superior performance, while other models, including linear models (LR and lasso) and neural network-based models, performed poorly in comparison to the model proposed in \cite {TurkishPaper}. 
The second-best performing model was the neural network. Only three hidden layers were added to the feed-forward neural network, as this configuration provided good performance with an acceptable training time. Adding more layers increased both training time and the risk of overfitting, while yielding only minimal performance improvement. Therefore, a trade-off between achieving good performance metrics and minimizing training time was considered. Moreover, the neural network's competent performance can be attributed to its ability to capture non-linear relationships between input features and manage high-dimensional input. However, one potential drawback of feed-forward networks is their interpretability. While decision trees are easily interpretable, neural networks are more challenging to understand due to their complex architecture. They are also computationally expensive to train, particularly when dealing with large datasets. \cite{FNN}. 
The third-best performing model, which achieved desirable MSE, RMSE, and  $R^2$ values, was the transformer model with self-attention and positional encoding. The self-attention mechanism allows the model to focus on the most relevant parts of the input (such as the readings of current and voltage over time) in our battery SOC prediction task. Furthermore, the positional encoding layer enables the model to capture temporal relationships between data points, which aids in analyzing the battery’s behaviour over time and detecting any trends that may be present  \cite{attention}.

Furthermore, the ablated transformer networks performed relatively poorly on this prediction task. This result emphasizes the essential need for having both the self-attention and positional encoding layers in the network. Their ability to capture relevant parts of the input and the temporal relationships between data points is essential for achieving competent performance in this SOC regression task.  
In addition, the performance of the linear and lasso regression models was unsatisfactory, as these models are primarily effective at capturing linear relationships between input variables and the target. The exploratory data analysis (EDA) performed on the dataset indicated that the relationships between the input features and the target are highly nonlinear. Consequently, the linear regression (LR) model was unable to capture these nonlinear relationships effectively, which was reflected in the poor performance metrics. Moreover, the nonlinear behaviour exhibited by the battery, particularly at high and low SOC levels, makes linear and lasso regression less effective. For instance, at these SOC levels, the battery's voltage does not change linearly with SOC, and similar nonlinear behavior is observed when the battery experiences significant temperature changes  \cite{nonlinearReg}. Therefore, more complex models like decision trees and neural networks are better suited for addressing such a problem. 

The performance metrics for all the models investigated in this task are summarized below in Table \ref{tab1}. 

\vspace {-0.3 cm}
\begin{table}[h]
  \begin{center}
   \caption{Performance metrics for all models}   
  \label{tab1}
\begin{tabular}{ |c|c|c|c|c| } 
 \hline
 ML Model & MSE & RMSE & R2 & MAE \\ 
 \hline
 Decision Tree & 	0.0000 &	0.0052 &	1.0000 &	0.0005 \\ 
 \hline
 Neural Network Regression &	0.0742 & 0.2724 & 0.9997 & 0.2048\\ 
 \hline
 Transformer with Self-Attention and Positional Encoding & 0.1228 &	0.3505 &  0.9995	& 0.7700 \\ 
 \hline
 Transformer with Self-Attention & 1.0757 &  1.0372 &	0.9954 & 0.7700 \\ 
 \hline
 Linear Regression	& 1.5534 &	1.2464 &	0.9933 &	0.8586\\
 \hline
 Lasso Regression	& 1.8052 &	1.3436 &	0.9922 &	0.9318\\ 
 \hline
Transformer with Positional Encoding &	2.9100 &	1.7059 &	0.9875 &	1.4487 \\ 
 \hline
\end{tabular}
\end{center}
\end{table}
\vspace{-40pt}
\subsubsection {Explanation of obtained performance metrics}

All models in this task achieved high  $R^2$ values, indicating that they explain a large proportion of the variability in the data and accurately capture underlying patterns  \cite{AnIntroductiontoStatisticalLearning}   \cite{hastie2009elements} \cite{ToexplainOrtopredict?}. The  $R^2$  metric measures the variance explained by the independent variables \cite{investopedia_rsquared}. However, while  $R^2$ is commonly used to evaluate regression models, it was not sufficient for this SOC prediction task alone. Therefore, additional metrics such as MSE, RMSE, and MAE were employed.

Reliance on $R^2$  alone can be misleading, as it does not account for overfitting or the model's ability to generalize to new data\cite{LimitationsOfR^2}\cite{ToexplainOrtopredict?}. Moreover, MSE might produce superficial results when dealing with large datasets. To address this, K-fold cross-validation was implemented to provide more reliable and insightful outcomes. 

Additionally, the decision tree, neural network, and transformer models with self-attention and positional encoding achieved desirable low MSE, RMSE, and MAE values. Low MSE and RMSE values indicate that the models are accurately predicting the target variable, capturing key patterns in the data, and generalizing well to unseen data. On the other hand, MAE measures the absolute difference between predicted and actual values. A low MAE value suggests that these differences are minimal, meaning the models can make precise predictions of the dependent variable \cite{AnIntroductiontoStatisticalLearning}.

Furthermore, some models, such as linear and lasso regression and the ablated transformer networks, exhibited contradictory performance with high MSE, RMSE, MAE, and $R^2$ values. This suggests overfitting to the training data and poor generalization to the test data. Evaluating multiple metrics and implementing K-fold cross-validation was crucial for assessing the models' generalization ability. This approach helped identify the most suitable models for the SOC prediction task. 

\section{Conclusion}

The use of data-driven approaches to predict the state of charge of lithium-ion batteries has great potential and can be used to replace conventional SOC estimation methods. In this project, several machine-learning approaches were implemented to predict the SOC of lithium-ion batteries in real driving cycles, and their overall performance was compared. The obtained results showed that the baseline decision tree model outperformed all other baseline models, as indicated by its relatively low MSE, RMSE, and MAE values, alongside a high $R^2$ value. This prominent performance of the decision tree model can be attributed to its robustness in handling real-world noisy data and its ability to capture nonlinear relationships between input features and the target variable. The model was also able to outperform the bagging random forest proposed in \cite{TurkishPaper} which was experimented on the same data scrutinized in this project. The second-best performing model was the neural network, which efficiently captured the nonlinear relationships between the input variables and the dependent variable. The more complex transformer model with self-attention and positional encoding ranked third in performance. The rest of the ablated transformer networks as well as the linear and lasso regression models performed relatively poorly in this regression task.  

\subsubsection {Model complexity: Is attention really all we need? }

Exploring various machine learning models for predicting the state of charge in lithium-ion batteries highlights the importance of starting with baseline models before progressing to more complex architectures, such as transformer networks. While transformers and neural networks have gained popularity for SOC regression tasks, they can lead to temporal information loss in datasets with time-dependent data due to the permutation-invariant nature of self-attention, even when positional encoding is applied \cite{doWeNeedTrans}. Therefore, simpler models should be considered initially before advancing to more complex architectures. The results in this paper emphasize the importance of this approach.

\subsubsection {Future work}
For future work, it would be interesting to investigate whether the methods and results obtained in this task can be generalized to other real-world charging and discharging data, as well as to different electric vehicles. Most machine learning models today offer black-box predictions that cannot be easily generalized to other battery chemistries \cite{ConventionalSOC}. Hence, incorporating domain knowledge is essential to achieve comprehensible and reliable predictions.  Moreover, high-throughput experimentation should be conducted to generate high-quality, real-world datasets. Reducing the reliance on large data storage devices will also be crucial for advancing data-driven methods for real-time battery modelling.  

\bibliography{paper}
\end{document}